\documentstyle[12pt]{article}

\catcode`\@=11

\def\@maketitle{\newpage   
 \null
 \vspace*{-1\headsep}      
 \vspace*{-1\headheight}
 \vspace*{-24pt}
 \begin{flushright}{\large       
   { \preprintno} \\ \@date}
 \end{flushright}
 \vskip \headsep     
 \vskip \headheight
 \bigskip
 \begin{center}            

{\LARGE \@title \par}
   \vskip 2em
   {\large
     \lineskip .5em
     \begin{tabular}[t]{c}\@author
     \end{tabular}\par}
   \vskip 1em
 \end{center}
 \par
 \vskip 1.5em}

\newcommand{\preprintno}{preprint number here}   

\def\abstract{\if@twocolumn
\section*{Abstract}
\else                         
\begin{center}
{\bf Abstract\vspace{-.5em}\vspace{0pt}}
\end{center}
\quotation
\fi}
\def\endabstract{\if@twocolumn\else\endquotation\fi}

\def\appendix{\par
    \setcounter{section}{0}
    \setcounter{subsection}{0}
    \renewcommand{\theequation}{\Alph{section}.\arabic{equation}}
    \setcounter{equation}{0}
}

\@addtoreset{equation}{section}
\def\theequation{\arabic{section}.\arabic{equation}}

\def\ps@columns{%
 \if@twocolumn
  \let\@mkboth\@gobbletwo
  \def\@oddhead{}\def\@evenhead{}
  \def\@oddfoot%
   {\rm\hfil\thepage\stepcounter{page}\hskip.5\textwidth\thepage\hfil}
  \let\@evenfoot\@oddfoot
 \else
 \ps@plain
 \fi
}

\catcode`\@=12

\makeatletter \input art10.sty \makeatother
\def\starttext{\twocolumn}

\typein{*** or hit [return] for landscape mode(11x8.5) ***}

\newcommand{\beq}{\begin{equation}}
\newcommand{\eeq}{\end{equation}}

\newcommand{\remove}[1]{}
\renewcommand{\theequation}{\thesection.\arabic{equation}}

\begin{document}

\title{Detecting the Technirho in Technicolor Models with Scalars}

\author{
        Christopher D. Carone\thanks {carone@huhepl.harvard.edu} \\
        Mitchell Golden\thanks {golden@physics.harvard.edu} \\
        Lyman Laboratory of Physics \\
        Harvard University \\
        Cambridge, MA 02138}
\date{\today}

\renewcommand{\preprintno}{HUTP-93/A034}

\begin{titlepage}

\maketitle

\def\thepage {}        

\begin{abstract}
We consider the detection of the technirho in technicolor
models that include a weak scalar doublet.  In these models, the
scalar develops a vacuum expectation value when the technifermions
condense, and fermion masses are generated through Yukawa couplings, as in
the standard model.  Unlike ordinary technicolor models, the
technicolor scale is generally smaller than the electroweak scale, and
the entire technicolor spectrum is shifted to lower energies.
In addition, the technifermions can develop large `current' masses
from the scalar's vev, so that the technicolor spectrum does not necessarily
look like a scaled-up version of ordinary QCD.   We discuss the
collider signatures for the technirho in these models in the limit that
the technirho is light.  Although we focus mostly on the production and
detection of the technirho in hadron colliders, we also discuss a peculiar
limit in which the technirho may be light enough to be seen at LEP II.
\end{abstract}

\end{titlepage}

\starttext 
\pagestyle{columns} 
\pagenumbering {arabic} 

\section {Introduction} \label {sec:intro}

Models that include both technicolor and scalars \cite{smodel,samuel}
may seem unorthodox to anyone who is familiar with the original
motivation for the introduction of technicolor.  Nevertheless, it has
been shown in a series of recent papers \cite{cgmodel,csmodel} that
minimal technicolor models with a weak scalar doublet can successfully
account for fermion masses without generating large flavor changing neutral
current effects, and without exceeding the experimental bounds on oblique
electroweak radiative corrections \cite{oblique}.
While technicolor with a massive scalar might arise in nature as the
low-energy effective theory for a strongly coupled ETC model \cite{chiv}, the
limit in which the scalar doublet is light is perhaps even more interesting.
In ref.~\cite{cgmodel} it was argued that technicolor with a massless scalar
is the {\it simplest} nontrivial extension of the standard model, with only
two new free parameters.  Like the Coleman-Weinberg model \cite{colwein}, the
model has no mass parameters, yet it does not suffer from any major
phenomenological problems.

In this paper, we consider the clearest
experimental signature of technicolor with scalars -- the detection of
both a Higgs boson and a light technirho.  In the limit in which the
technicolor scale is low, the Higgs boson in technicolor models with
scalars is indistinguishable from its counterpart in the standard
model \cite{cgmodel}, and we need not consider it any further; Higgs
boson detection has already been covered extensively in the
literature \cite{higgs}.  We will instead focus on the production and
detection of the light technirho at $pp$ colliders.  The technirho in
models with scalars is different from its counterpart in ordinary technicolor
models, not only because it can be much lighter than the electroweak scale,
but also because the technifermions have significant current masses relative
to the technicolor scale in the same limit.   Thus, technirho
phenomenology cannot be obtained by studying the $\rho$ meson in QCD
and applying simple scaling arguments.

The models that we will consider in this paper have a
minimal $SU(N)$ technicolor sector, consisting of two techniflavors, $p$ and
$m$.
The technifermions transform as a left-handed doublet and two right-handed
singlets under $SU(2)_W$:
\beq
\Upsilon_L=\left(\begin{array}{c} p \\ m \end{array}
\right)_L, ~~ p_R, ~~ m_R ~~.
\eeq
The hypercharges are $Y(\Upsilon)=0$, $Y(p_R)=1/2$, and
$Y(m_R)=-1/2$, which cancels gauge anomalies.  The technifermions
transform as singlets under color and as fundamentals under the $SU(N)$
technicolor group.  The ordinary fermions are technicolor singlets,
but otherwise have the same quantum numbers as in the standard
model.

The feature that distinguishes this model from miminal
technicolor alone is the existence of a scalar field $\phi$ which transforms
as a weak doublet with hypercharge $1/2$
\beq
\phi = \left(\begin{array}{c} \phi^+ \\ \phi^0\end{array} \right)
\eeq
Like the Higgs doublet of the standard model, the scalar has Yukawa
couplings to the ordinary fermions.  In addition, it also couples to
the technifermions
\beq
{\cal L}_{\phi T} = \overline{\Upsilon}_L \tilde{\phi} h_+ p_R
+\overline{\Upsilon}_L \phi h_- m_R + \mbox{h.c.}
\label{eq:tfcoup}
\eeq
where $h_{\pm}$ are Yukawa couplings, and $\tilde{\phi}^i = \epsilon^{ij}
\phi^\dagger_j$.  When technicolor becomes strong
the technifermions form a condensate which breaks the electroweak
symmetries to $U(1)_{EM}$.  In addition, the condensate generates a linear
term in the scalar potential, through the Yukawa couplings in
(\ref{eq:tfcoup}).  As a result, the scalar develops a vacuum expectation
value, and fermion masses are generated through the Yukawa couplings, in
the usual way.  This mechanism was explored in some detail in
ref.~\cite{cgmodel}.

The technirho in this model is the isotriplet vector bound state of
the technifermions.   However, it need not look anything
like the ordinary $\rho$ in QCD when the Yukawa couplings in
(\ref{eq:tfcoup}) are large. The same mechanism responsible for fermion
mass generation gives the technifermions current masses as well.
We define the parameters $h$ and $\delta$ by
\beq
h\equiv\frac{h_++h_-}{2} ~~~
\delta\equiv \frac{h_+-h_-}{h_++h_-}~~.
\eeq
We will be interested in studying the model in the limit of large $h$, since
this corresponds roughly to the limit in which the technicolor scale is low
and the technirho is light.  In this case, the technirho is more similar to a
bound state of heavy flavors in QCD, than it is to the ordinary $\rho$ meson.
We will study two limits: $\delta=0$ and $\delta=1$.  If $\delta=0$, the $p$
and $m$ technifermions have equal current masses, while if $\delta=1$, the $p$
is heavy while the $m$ is massless.  In the latter case, we will argue that
the vector bound state of the massless technifermions may be light enough to
be seen at LEP II.  Given the uncertainties in our analysis, we will make no
attempt to sample the full range in $\delta$, nor the full parameter space of
the model.  Rather, we will attempt to draw some qualitative conclusions by
studying technirho phenomenology in these two interesting limits.

\section {Operators} \label {sec:op}

Our goal in this section is to extend the chiral lagrangian
analysis of refs. \cite{cgmodel,csmodel} to include the operators
that are responsible for technirho production and decay. We begin by
catagorizing the particles by their transformation properties
under the $SU(2)_L \times SU(2)_R$ symmetry of the technicolor
sector.  We adopt the conventional nonlinear representation
of the technipion fields
\beq
\Sigma = \exp\left(2 \, i \, \Pi/f\right)\,, \quad\quad
\Sigma \rightarrow L\Sigma R^\dagger\,.
\label{eq:Sigma}
\eeq
where
\beq
\Pi = \frac{1}{\sqrt{2}}\left[
\begin{array}{cc} \pi^0/\sqrt{2} & \pi^+ \\
          \pi^-     & -\pi^0/\sqrt{2}
\end{array}\right]\,.
\label{eq:pimatrix}
\eeq
The fields in $\Pi$ represent the pseudoscalar bound states of the
technifermions $p$ and $m$, and $f$ is the associated decay constant.
In addition, we define the field $\xi$ by
\beq
\xi \xi = \Sigma
\label{eq:xidef}
\eeq
where $\xi$ transforms as
\beq
\xi \rightarrow L \xi U^\dagger = U \xi R^\dagger
\label{eq:xiprop}
\eeq
under the chiral symmetry.  Note that eq. (\ref{eq:xiprop}) defines
the matrix $U$ implicitly in terms of the technipion fields.  Using
$\xi$ and $\xi^\dagger$, we may construct operators that transform simply
under the local transformation defined by $U$, and then build all the
possible independent invariants. This will give us the appropriate
low-energy effective theory.

To determine the physical scalar states in the low-energy
theory, we must consider the mixing of the technipions in
(\ref{eq:pimatrix}) with the scalar $\phi$.  If we write $\phi$ in the
matrix form
\beq
\Phi=\left[\begin{array}{cc} \overline{\phi^0} & \phi^+ \\
                    -\phi^- & \phi^0 \end{array}\right]
\eeq
then the kinetic energy terms are given by
\beq
{\cal L}_{K.E.} = \frac{1}{2} \mbox{Tr }({\cal D}_\mu\Phi^\dagger
{\cal D}^\mu \Phi) + \frac{f^2}{4}\mbox{Tr }({\cal D}_\mu \Sigma^\dagger
{\cal D}^\mu \Sigma)
\label{eq:ke}
\eeq
where ${\cal D}^\mu$ is the $SU(2)_L\times SU(2)_R$ covariant derivative
\beq
{\cal D}^\mu\Sigma = \partial^\mu\Sigma - i g W^\mu_a\frac{\sigma^a}{2} \Sigma
+ i g' B^\mu \Sigma \frac{\sigma^3}{2} \,\,\,.
\eeq
To isolate the physical states, we reparametrize $\Phi$ in terms of an
isoscalar field $\sigma$ and an isotriplet $\Pi'$,
\beq
\Phi = \frac{(\sigma + f')}{\sqrt{2}} \Sigma '
\eeq
\beq
\Sigma' = \exp (2 i \Pi' / f')
\eeq
where $f'$ is the scalar vev.  It is now easy to show from
(\ref{eq:ke}) that the pions in the linear combination
\beq
\pi_a = \frac{f \Pi + f' \Pi '}{\sqrt{f^2+{f'}^2}}
\eeq
become the longitudinal components of the $W$ and $Z$
in unitary gauge, while the orthogonal combination
\beq
\pi_p = \frac{-f' \Pi + f  \Pi '}{\sqrt{f^2+{f'}^2}}
\label{eq:pp}
\eeq
remain as physical scalars in the low-energy theory.  In addition,
the requirement that we obtain the correct gauge boson masses from
(\ref{eq:ke}) gives us the constraint
\beq
f^2 + {f'}^2 = v^2
\eeq
where $v\approx 250$ GeV is the electroweak scale.

In this paper, we consider the regime in which $f' \sim v$, $f \ll v$, and the
technirho is light.  As one can see from (\ref{eq:pp}), the physical
pions $\pi_p$ become almost pure technipion states, so it will
make little difference whether we include the small mixing with the
fundamental scalar in arriving at our numerical estimates.  The condition that
there is no linear term in the Higgs potential after symmetry breaking gives
a second constraint involving $f$ and $f'$, which allows us to solve for
each in terms of the free parameters of the model\cite{cgmodel,csmodel}.
Furthermore, when $f$ is small, the current masses of the technifermions
become sizeable compared with the scale $4\pi f$, which means that the
technirho becomes very narrow, as we will in the next section.

Of course, this particular region of parameter space is not generic.  A theory
with $f$ and $f'$ not much smaller than $v$ will have both a Higgs boson and a
technirho, each with more or less their usual properties.  If it's heavy
enough, the Higgs boson decays to pairs of gauge bosons, otherwise to the
heaviest fermion allowed.  The technirho is heavy and decays to pairs of
longitudinal gauge bosons.  The processes of Higgs boson production and decay
via gauge boson scattering are reduced by a factor of $(f'/v)^4$.  The other
corner of parameter space has $f \sim v$ and $f' \ll v$.  This is
phenomenologically problematic because the quark masses are given by
$f'$ times a Yukawa coupling matrix; if $f'$ is too small, the Yukawa
couplings become large and either the flavor changing neutral current effects
exceed the experimental limits, or the physics of the top quark becomes
non-perturbative.

In addition to the terms that we have considered, we can also construct
interactions that explicitly involve the scalar $\Phi$.  In matrix form, the
scalar's coupling to the technifermions can be written
\beq
\overline{\Upsilon}_L \Phi H \Upsilon_R + \mbox{h.c.}
\label{eq:mcoup}
\eeq
where $H = \mbox{diag}(h_+\,,\,h_-)$ is the matrix of the technifermion
Yukawa couplings.  If $\Phi H$ were to transform as
\beq
\Phi H \rightarrow L (\Phi H) R^\dagger
\label{eq:spur}
\eeq
then  (\ref{eq:mcoup}) would be left invariant.  Thus, the appropriate
way to include $\Phi H$ in the chiral expansion is to assume
the transformation law (\ref{eq:spur}), and build all the
possible invariants.

To incorporate the technirho in our chiral lagrangian, we choose the
technivector mesons to transform as a triplet
\beq
M^\mu \rightarrow U M^\mu U^\dagger
\label{eq:vmm}
\eeq
where
\beq
M = \left[\begin{array}{cc} \rho^0/2 +\omega/2& \rho^+/\sqrt{2} \\
                             \rho^-/\sqrt{2} & -\rho^0/2 +\omega/2
             \end{array}\right]
\eeq
If we were to choose a different transformation law for $M$,
we would obtain an equivalent effective theory \cite{wimpt}.  Note that by
placing the isosinglet and isotriplet states together in a single multiplet,
we have imposed a ``quartet'' symmetry, which is explicitly broken by terms in
the lagrangian proportional to $\mbox{Tr } M^\mu$;  However, these
symmetry breaking terms involve at least two traces over flavor indices, and
are therefore suppressed in the large $N$ technicolor limit; for this
reason, we assume these terms are small and simply ignore them.

Since we will want to construct operators that transform properly
under parity, we define the vector and axial vector fields
\beq
V^\mu = -\frac{i}{2} (\xi^\dagger \partial^\mu _L \xi
+ \xi \partial^\mu _R \xi^\dagger)
\label{eq:vdef}
\eeq
\beq
A^\mu = \frac{i}{2} (\xi^\dagger \partial^\mu _L \xi
- \xi \partial^\mu _R \xi^\dagger)
\label{eq:adef}
\eeq
where $\partial^\mu _{L(R)}$ is the $SU(2)_{L (R)}$
covariant derivative,
\beq
\partial^\mu _L = \partial^\mu + i g W^\mu _a \frac{\sigma_a}{2}
\label{eq:lcov}
\eeq
\beq
\partial^\mu _R = \partial^\mu + i g B^\mu  \frac{\sigma_3}{2}
\label{eq:rcov}
\eeq
Notice that $A^\mu$ transforms as an isotriplet, while
$V^\mu$ transforms as a gauge field.  Thus, we can define a covariant
derivative
\beq
D^\mu \rho^\nu = \partial^\mu \rho^\nu + i [V^\mu , \rho^\nu]
\eeq
which transforms homogeneously under the local transformation
\beq
D^\mu \rho^\nu \rightarrow U (D^\mu \rho^\nu) U^\dagger
\eeq
Finally, we define the parity even and odd combination of the
electroweak gauge field strength tensors
\beq
\tilde{F}^{\mu\nu}_{\pm} \equiv \xi^\dagger g W^{\mu\nu}_a \frac{\sigma_a}{2}
\xi \pm
\xi g' B^{\mu\nu} \frac{\sigma_3}{2} \xi^\dagger
\label{eq:ftwid}
\eeq
Notice that $\tilde{F}_{\pm}^{\mu\nu}$ goes into
$\pm\tilde{F}_{\pm\,\mu\nu}$ under the parity transformation:
$\xi \leftrightarrow \xi^\dagger$, $\partial^\mu
\leftrightarrow \partial_\mu$, and $g W^\mu \leftrightarrow g' B_\mu$.
It is also useful to note that the objects we have
constructed transform under charge conjugation
as: $M_\mu \rightarrow - M_\mu^*$, $A_\mu
\rightarrow A_\mu^*$, and $F_{\mu\nu} \rightarrow - F_{\mu\nu}^*$.

Now we are in a position to write down the lowest order
technirho operators.  Assuming that the technicolor sector
is invariant under parity and charge conjugation, the complete isospin
invariant set is:
\beq
\frac{c_1}{4 \pi} \mbox{Tr } \tilde{F}^{\mu\nu}_+ D_\mu M_\nu
\label{eq:c1}
\eeq
\beq
\frac{c_2}{4 \pi} \epsilon_{\alpha\mu\nu\beta} \mbox{Tr }
\tilde{F}^{\mu\nu}_- D^\alpha M^\beta
\label{eq:c2}
\eeq
\beq
\frac{c_3}{4 \pi} \epsilon_{\alpha\mu\nu\beta} \mbox{Tr }
\left\{A^\alpha\,,\,\tilde{F}^{\mu\nu}_+ \right\}M^\beta
\label{eq:c3}
\eeq
\beq
\frac{c_4}{4 \pi} \mbox{Tr }
\left\{ A_\mu \,,\, \tilde{F}^{\mu\nu}_-\right\} M_\nu
\label{eq:c4}
\eeq
\beq
i \frac{c_5}{4 \pi} \mbox{Tr } \left[ A^\mu, A^\nu \right] D_\mu
M_\nu
\label{eq:c5}
\eeq
\beq
i \frac{c_6}{4 \pi} \mbox{Tr } \left[ A^\mu\,,\, D_\sigma A_\mu\right]
M^\sigma
\label{eq:c6}
\eeq
\beq
i \frac{c_7}{4 \pi} \mbox{Tr } \left[ A^\mu \,,\, D^\nu A_\nu\right]
M_\mu
\label{eq:c7}
\eeq
\beq
i \frac{c_8}{4 \pi} \epsilon_{\mu\nu\alpha\beta}
\mbox{Tr } A^\mu A^\nu A^\alpha M^\beta
\label{eq:c8}
\eeq
One can show with the help of the identities
\beq
D^\mu A^\nu - D^\nu A^\mu + \frac{1}{2} \tilde{F}^{\mu\nu}_- = 0
\eeq
and
\beq
[D^\mu\,,\, D^\nu] - [A^\mu\,,\,A^\nu] - \frac{i}{2} \tilde{F}^{\mu\nu}_+ = 0
\eeq
that any other operator of the same order can be expressed as
a linear combination of the operators given in this list.

There are in addition operators that contain the isospin breaking spurion
combination $(\Phi H)$.  If we define the parity even and odd combinations
\beq
S = \frac{1}{2} \left(\xi^\dagger \Phi H \xi^\dagger + \xi H \Phi^\dagger
\xi \right)
\eeq
\beq
P = \frac{i}{2} \left(\xi^\dagger \Phi H \xi^\dagger - \xi H \Phi^\dagger
\xi \right)
\eeq
then the lowest order term involving one factor of $H$ is given by
\beq
i c_{9} f \mbox{Tr } \left[A^\mu , P \right] M_\mu
\label{eq:c11}
\eeq
We found that its contributions to the technirho decays of interest are
no larger than the contributions from the other operators on our list.

Operators that include more derivatives (either in the form of additional
$A^\mu$s or $D^\mu$s) or more $\Phi H$s are suppressed relative to the terms
shown by powers of some scale, conventionally denoted $\Lambda_{TC}$.  If we
assume that $\Lambda_{TC} \gg m_\rho$, then we need not include any terms with
more derivatives.  This is not really a very good assumption.  However, we
will take our cue from QCD, in which the predictions made in this way
seem to work reasonably well.

The first operator, (\ref{eq:c1}), is responsible for vector dominance; it
contains $\rho^0$-$\gamma$, $\rho$-$Z^0$ and $\rho^\mp$-$W^{\pm}$ couplings.
We assume that the dimensionless coefficient $c_1$ is the same as it is for
the physical $\rho(770)$.  There is no easy way to justify this assumption,
except to note that an approximate equality holds for the comparible
coefficients of the $\rho(770)$ and $\phi(1020)$ in QCD.  With this
assumption, we use this operator to determine the technirho production cross
section at $pp$ colliders, and we show, in table I, the number of technirhos
produced at the LHC per year through vector dominance, assuming a center of
mass energy of 14 TeV and a luminosity of $10^{33}$~cm$^{-2}$~s$^{-1}$. The
total cross sections were obtained using the EHLQ set II structure functions
\cite{ehlq}, and include a rapidity cut ($y<2$) to eliminate the technirhos
produced in the beam direction.

\begin{center}
\begin{tabular}{lcccc}
\multicolumn{5}{c}{{\bf Table I}} \\
Mass (GeV)         &  &\multicolumn{3}{c}{Number/LHC-Year}  \\ \hline\hline
            &  &$\rho^+$        &  $\rho^0$     &  $\rho^-$  \\ \hline
200         &  &$1\times 10^5$  & $5\times 10^5$  & $1\times 10^5$ \\
250         &  &$7\times 10^4$  & $3\times 10^5$  & $5\times 10^4$ \\
300         &  &$4\times 10^4$  & $1\times 10^5$  & $3\times 10^4$ \\
350         &  &$2\times 10^4$  & $1\times 10^5$  & $2\times 10^4$ \\
\hline\hline
\end{tabular}
\end{center}
\begin{center}
1 LHC-Year = $10^{40}$ cm$^{-2}$ at $\sqrt{s}$ = 14 TeV.
\end{center}

In Table I we have set $\delta=0$, so the different charge states are
degenerate in mass.   When $\delta=1$ the production of the lightest
vector meson state, the neutral $m\overline{m}$ bound state,
is simply $1/2$ times the $\rho^0$ results shown in Table I, as we will
explain in section \ref{sec:done}.  Note that technirhos are also produced
by gauge boson fusion \cite{chanow}, but we found production through this
mechanism to be negligible in comparison to the results presented above.

In sections \ref{sec:dzero} and \ref{sec:done}, we will focus on the
remaining operators in our list to determine the technirho branching
fractions, and to isolate the most promising modes for detection.

\section{The $\delta=0$ Limit} \label{sec:dzero}

To compute the technirho branching fractions, we need to estimate the
masses of the technirho and of the physical scalar states in the
low-energy theory. In a minimal technicolor model, we would arrive at
the technirho mass by `scaling up' the mass of the ordinary $\rho$
in QCD
\beq
m_{\rho_T} \approx \left(\frac{3}{N}\right)^\frac{1}{2}
\left(\frac{f}{93 \mbox{MeV}}\right) m_\rho
{}~~~~\mbox{(minimal technicolor)}
\label{eq:stupid}
\eeq
However, (\ref{eq:stupid}) is not appropriate to the models of interest
to us, because the technifermions can develop significant current masses
from the scalar vev.  In ref. \cite{cgmodel} it
was shown that the region of parameter space in which the technirho is
light ($f\approx 25$ GeV) is also the region in which the technifermions
have current masses that are of the same order as the rho mass estimated from
(\ref{eq:stupid}), that is $hf'\approx 200$ GeV.  Thus, we would expect that
the low-lying vector meson state is more similar to the $\phi$ meson in
QCD (an $s \overline{s}$ bound state) than it is to the ordinary $\rho$.
To take the technifermion current masses into account, we use a slightly
more complicated algorithm for estimating the technirho mass: First we plot
the masses of vector meson states in QCD, as a function of the quark
current mass. As shown in Fig. \ref{fig:one}, the data points correspond
to the $\rho$, the $\phi$, the $J/\psi$, and the $\Upsilon (1S)$, in order of
increasing quark mass, $m_q$.  Then we fit these points with a straight
line, so that we can determine the vector meson mass $M_V$ in QCD for
intermediate values of the quark current mass.  While the actual dependence
of the technirho mass on the quark mass is probably not linear for
small $m_q$, this will not effect our qualitative results.  The value
of $m_q$ that is relevant to us is given by
\beq
m_q = \frac{m_{TF}}{f} f_\pi
\eeq
where $m_{TF}$ is the technifermion current mass at the desired point in the
parameter space of the technicolor model, and $f_\pi = 93$ MeV.  Finally, we
take the corresponding vector meson mass and naively scale it up as in
(\ref{eq:stupid}).

To compute accurately the decay rates of the technirho, we would need an
estimate of the physical pion mass.  Unfortunately the same large current
quark mass which made it difficult to estimate the technirho mass makes
estimates of the pion mass from the chiral lagrangian unreliable.  Instead,
we make a simple qualitative argument.  As we increase $h$, $f \to 0$.
Therefore, the physical triplet pions become mostly pure technipion states.
We expect the splitting of $m_{\rho_T}$
and $m_{\pi_p}$ to become small in this limit for the same reason that the
splitting of the $J/\psi$ and the $\eta_c$ masses is smaller than that of
the $\rho$ and the $\pi$ in QCD:  the hyperfine splittings are smaller for
mesons containing heavy quarks.  Thus, in the limit of our model in which the
technicolor scale is low, and in which the technifermion current masses
are significant, we believe that it is reasonable to assume that the
inequality
\beq
m_{\rho_T} < 2 m_{\pi_p}
\eeq
is always satisfied.  This is crucial phenomenologically because the
decay $\rho \rightarrow \pi_p \pi_p$ is kinematically forbidden.

With the two-pion decay mode ruled out, we can proceed
to investigate the remaining possibilities.  We will focus on the
decay modes of the $\rho_T^0$, which will be produced roughly 70\% of
the time.  In addition to having a smaller production rate, the charged
technirho will decay primarily to final states that do not allow a full
reconstruction of the event. Thus, we believe that the decays of the neutral
state will provide a more promising experimental signature.  As a
consequence of unbroken custodial isospin, we can immediately rule out
$\rho^0_T$ decays into the following final
states: $\gamma \gamma$, $\gamma Z^0$, $Z^0 Z^0$, $\gamma
\pi_p^0$, and $Z^0 \pi_p^0$.  The modes that remain are:  $W^\pm \pi_p^\mp$,
$W^+W^-$, $l^+l^-$, $q\overline{q}$, $\eta '_T \gamma$, and $\eta '_T Z^0$,
where $\eta'_T$ is the isoscalar technifermion bound state.  Although we have
not yet discussed the $\eta'_T$, we will have more to say about it shortly.
To estimate the decays widths, we evaluate the operators listed in section
\ref{sec:op}. When two or more operators contribute to a specific decay
mode, we evaluate only the largest of them to obtain an order of magnitude
estimate. This approach is reasonable -- we don't know the relative signs
of the various operators, but we don't expect miraculous cancelations either.
In addition, we set the coefficients $c_j$ to their values in QCD, for the
operators that are relevant to our numerical analysis.  We find
$c_1 \approx c_3 \approx 2$ and $c_5 \approx 4$, consistent with naive
dimensional analysis \cite{manran}.  Again, there is no good way to justify
naive dimensional analysis for quarks this heavy, except that it seems to
work reasonably for the $\phi(1020)$.  Thus, we believe our choice of
coefficients is a reasonable one.  The explicit forms of the decay widths
are provided in the appendix.  For a 250 GeV $\rho_T^0$, the decay modes
to gauge bosons and fermions alone are summarized in Table II.

\begin{center}
\begin{tabular}{lcc}
\multicolumn{3}{c}{{\bf Table II}} \\
\multicolumn{3}{c}{$M_{\rho_T}=250$ GeV} \\
$\rho_T^0$ Decay Mode &  & Decay Width (GeV)\\ \hline\hline
$W^+W^-$    &  & $6.1\times 10^{-3}$ \\
$u\overline{u}$ & & $3.7\times 10^{-3}$ \\
$d\overline{d}$ & & $2.5\times 10^{-3}$ \\
$e^+e^-$    &  & $1.8\times 10^{-3}$  \\
$\nu_e\overline{\nu}_e$ & & $5.2\times 10^{-4}$ \\
\hline\hline
\end{tabular}
\end{center}

The decay to $\pi_p^\pm W^\mp$ depends on the mass of
the the $\pi_p$.  In Fig.\ref{fig:piw}, we plot the decay width to
$\pi_p^+ W^-$ for a $250$ GeV technirho, as a function of
$m_{\pi_p}/m_{\rho_T}$.  We assume that $m_{\rho_T}<2m_{\pi_p}$ and
$f \approx 25$ GeV.  In the region of the plot in which the decay is
kinematically allowed, it dominates over the modes listed in Table II by
one to two orders of magnitude.  However, before we can draw any
qualitative conclusions, we need also to consider the decays involving
the $\eta'_T$.  To derive the form of its interactions, we will include
the $\eta'_T$ in our chiral lagrangian as an axial $U(1)$ pseudogoldstone
boson. We will set the associated decay constant $f_{\eta'}$ is equal to
$f$, as one would expect in the large $N_{TC}$ limit.
In this case, we can rewrite the pion matrix as
\beq
\Pi = \frac{1}{\sqrt{2}}\left[
\begin{array}{cc} \pi^0/\sqrt{2} + \eta'/\sqrt{2} & \pi^+ \\
          \pi^-     & -\pi^0/\sqrt{2} + \eta'/\sqrt{2}
\end{array}\right]\,.
\eeq
and find the desired interactions by evaluating the operators of
section \ref{sec:op}.  Of course, the axial $U(1)$ symmetry
is strongly broken by the anomaly, so the $\eta '_T$ is far from massless.
In QCD, the $\eta '$ is heavier than the $\rho$, but this need not be the
case in our technicolor model, because our technirho is built out of
technifermions with large current masses.  If the technirho is comparable to a
scaled up version of the $\phi$ meson in QCD, then it may be heavy enough for
the decay to $\eta '_T \gamma$ to proceed, but perhaps not heavy enough to
allow the $\eta' Z$ mode.  While we can attempt to estimate the $\eta'_T$ mass
from the $\eta'$ mass in QCD by means of scaling arguments, the uncertainties
involved will not allow us in the end to distinguish between these
kinematical possibilities.  Therefore, we will consider a range in
$m_{\eta'_T}$, subject to the constraint $m_{\rho_T}<2 m_{\eta'_T}$  In Fig.
\ref{fig:etap} we plot the decay width of a $250$ GeV technirho to $\eta'_T
\gamma$ and $\eta'_T Z$ as a function of $m_{\eta'_T}/m_{\rho_T}$.  From the
results in Table II, Fig. \ref{fig:piw}, and Fig.\ref{fig:etap}, we can draw
the following qualitative conclusions:

{\it (i)}   It is possible that both the $\pi_p$ and $\eta'_T$
are heavy enough so that the decays $\rho_T\rightarrow W^+ \pi^-_p$,
$\rho_T \rightarrow \eta'_T \gamma$, and $\rho \rightarrow \eta'_T Z^0$
are either forbidden or kinematically suppressed.  In this case, the
technirho decays primarily to the states given in Table II, (as well
as to the corresponding states in the other two generations), with
a total width of approximately $3\times 10^{-2}$ GeV.  If this is the
case, then the simplest method of detecting the $\rho_T^0$ would be through
its decays to $e^+e^-$ and $\mu^+\mu^-$, which occur roughly 11\% of
the time.  For our 250 GeV technirho, this roughly corresponds to
$3\times 10^4$ events per year at the LHC.

{\it (ii)}   If the $\eta'_T$ is light, then the decay
$\rho_T \rightarrow \eta'_T \gamma$  will be between one and
two orders of magnitude larger than the decay modes listed in
Table~II.  If the $\pi_p$ is also light, the $W^+ \pi_p^-$ mode
can compete with $\eta'_T \gamma$.  In either case, the branching
fraction to $e^+e^-$ and $\mu^+\mu^-$ is of order 1\%, corresponding
to $3\times 10^3$ events per year at the LHC.  Note that the
$\eta'_T \gamma$ mode would be easy to identify at an $e^+e^-$ collider
because of the isolated, monochromatic photon in the final state.

The general point that the reader should extract from this
example is that the light technirho in our
model is extremely narrow in the limit in which strong $\pi_p \pi_p$ modes
are kinematically forbidden.  As a result, the branching fraction
to $e^+e^-$ and $\mu^+\mu^-$ is large enough to allow detection
of the technirho, regardless of whether the $\eta'_T \gamma$
or $W^+ \pi_p^-$ modes dominate.  Using the results quoted in
reference \cite{ehlq} and assuming a detector resolution of 1\%, we find
that the number of background $l^+l^-$ events with an invariant mass of
$250$ GeV will be of order $10^2$, far smaller than the number of events
that we obtained in either case above.

\section{The $\delta=1$ Limit} \label{sec:done}

In this section, we consider the case where $\delta=1$, the limit of maximal
custodial isospin violation in the technicolor sector.  In
ref.~\cite{cgmodel}, it was shown that $\delta$ could be set equal to $1$
throughout most of the parameter space of the massless scalar model, without
generating a large contribution to the $T$ parameter.  Qualitatively,
the largest non-standard contribution to $T$ is from the $p$-$m$ mass
splitting, which in the limit of interest to us, is of the same order as the
the top-bottom mass splitting in the standard model.  Thus, we don't expect
drastic conflict with the experimental constraints.  Of course, if the
models of interest to us are realized in nature, $\delta$ could fall
anywhere in the range from $0$ to $1$.  The $\delta=1$ limit is particularly
interesting because one of the technifermions gets no contribution to its
mass from the scalar vev. It is clear from the technifermion current mass
matrix
\beq
\frac{f'}{\sqrt{2}}
\overline{\Upsilon_L} \left( \begin{array}{cc} h(1+\delta) & 0 \\
                                        0       & h(1-\delta)
            \end{array}\right) \Upsilon_R + \mbox{h.c.}
\eeq
that the $m$ technifermion becomes massless when $\delta=1$, while
the $p$ technifermion is heavy.  Therefore, we will integrate out
the heavy techniflavor, and work with an effective theory that is
a scaled up version of QCD with a single flavor.  In the absence
of a current mass contribution, the low-lying $m\overline{m}$
bound states will be lighter than the low-lying states
considered in section \ref{sec:dzero}.  Our hope {\it a priori} is that
the technirho may be light enough in this limit to be within reach of the
LEP II.

We first need to determine the physical vector and pseudoscalar
states in the low-energy theory, and to arrive at reasonable estimates of
their masses.  Since the technirho is now the low-lying vector
$m\overline{m}$ bound state, we can write it as a linear combination of
the $\delta=0$ fields presented in section \ref{sec:op}
\beq
\tilde{\rho}^0  = \frac{1}{\sqrt{2}}(- \rho^0 + \omega )
\label{eq:rhomix}
\eeq
where $\tilde{\rho}$ represents the mass eigenstate in
the $\delta=1$ limit.  The vector meson matrix (\ref{eq:vmm}) now takes
the form
\beq
M^\mu = \frac{1}{\sqrt{2}}\left[ \begin{array}{cc}    0  & 0 \\
                                    0  &   \tilde{\rho}^0
                \end{array} \right]
\label{eq:nrho}
\eeq
Since the $\tilde{\rho}$ is a bound state of massless technifermions,
we will estimate its mass by naive scaling from QCD:
\beq
m_{\tilde{\rho}} = \left(\frac{3}{N}\right)^\frac{1}{2}
\left(\frac{f}{93 \mbox{MeV}} \right)m_\rho
\label{eq:rtm}
\eeq
There is an assumption implicit in the use of this equation.  In QCD there are
three light quarks, and the $\rho$ and $\omega$ are approximately degenerate.
The $u \bar u$ combination, eqn. (\ref{eq:rhomix}), is therefore an
approximate mass eigenstate of QCD.  If we were to make the $d$ and $s$ quark
heavy, the assumption goes, the mass of $u \bar u$ combination would be
unaffected.  Only the change of the number of colors affects its mass.

In the massless
scalar model, it was shown that $f$ could be as low as $25$ GeV without
entering the excluded region of the model's parameter space.
This corresponds to a $\tilde{\rho}$ mass of $180$ GeV for $SU(4)$
technicolor, and $147$ GeV for $SU(6)$ technicolor. In both cases, the
$\tilde{\rho}$ would be within the reach of LEP II. \footnote{For
larger technicolor groups, we would exceed the constraints on the
$S$ parameter.}

The scalar $m\overline{m}$ bound state, which we will call $\tilde{\pi}^0$,
can also be written as a linear combination of $\delta=0$ fields,
\beq
\tilde{\pi}^0 = \frac{1}{\sqrt{2}} (- \pi^0 + \eta'_T)
\eeq
The pion matrix of section \ref{sec:dzero} then takes the form
\beq
\Pi = \frac{1}{\sqrt{2}}\left(\begin{array}{cc} 0 & 0 \\
                              0 & \tilde{\pi}^0
            \end{array}\right)
\eeq

In the $\delta=0$ limit, we argued that $m_{\rho_T} < 2 m_{\pi_p}$
in analogy to QCD, where the hyperfine splittings are small for mesons
containing heavy quarks.  This was important in that it allowed us to
eliminate the two-pion decay modes kinematically.  In the present case,
the two-pion decays are also irrelevant.  The decay to charged
pions is kinematically forbidden because the charged scalar states each
contain the heavy $p$ technifermion.  The remaining possibility,
$\tilde{\rho}\rightarrow \tilde{\pi}^0\tilde{\pi}^0$, is forbidden
by the Bose statistics of the particles in the final state.  Again,
we are left only with the weak decays to final states involving zero or one
scalar.

Now we can consider the remaining decay modes.  Since custodial isospin
is maximally violated in the $\delta=1$ limit, the $\gamma Z^0$ and
$Z^0 Z^0$ decay modes do occur, in addition to $l^+ l^-$, $q\overline{q}$,
$W^+W^-$, $\tilde{\pi}^0 \gamma$ and $\tilde{\pi}^0 Z^0$. We find that
the decay widths to fermion-antifermion pairs, and to $W^+W^-$ are simply
reduced by a factor of 2 compared to the same widths in the $\delta=0$
limit.  On the other hand, the $\tilde{\pi}^0 \gamma$ and
$\tilde{\pi} Z^0$ decay widths can be obtained from the $\eta'_T$ decay
widths of the $\delta=0$ limit by making the replacement
$m_{\eta'_T}\rightarrow m_{\tilde{\pi}}$.  The only modes that we haven't
already computed are those to $Z^0\gamma$ and $Z^0 Z^0$, and these are
provided in the appendix.  For a $180$ GeV technirho, the results for the
decays to fermions and gauge bosons alone are summarized in Table III.

\begin{center}
\begin{tabular}{lcc}
\multicolumn{3}{c}{{\bf Table III}} \\
\multicolumn{3}{c}{$M_{\tilde{\rho}}=180$ GeV} \\
Decay Mode  &  & Decay Width (GeV)   \\ \hline\hline
$u\overline{u}$ & & $1.5\times 10^{-3}$ \\
$d\overline{d}$ & & $1.1\times 10^{-3}$ \\
$Z^0 \gamma$    & & $7.2\times 10^{-4}$ \\
$e^+e^-$    &  & $6.8\times 10^{-4}$  \\
$\nu_e\overline{\nu}_e$ & & $2.6\times 10^{-4}$ \\
$W^+W^-$    &  & $ 1.6 \times 10^{-4}$  \\
$Z^0Z^0$    &  &         -            \\
\hline\hline
\end{tabular}
\end{center}

We plot the decay width to $\tilde{\pi}^0 \gamma$ in Fig. \ref{fig:four}
as a function of $m_{\tilde{\pi}}/m_{\tilde{\rho}}$.  The decay to
$\tilde{\pi} \gamma$ is the favored mode by one to two orders
of magnitude over the decays in Table III, for a $\tilde{\pi}$ that is
light.  Our argument that the leptonic modes could still be detected at $pp$
colliders carries over to the $\delta=1$ limit.  However, since the
$\tilde{\rho}^0$ is so light, it would make sense to look instead for the
decays to $\tilde{\pi}\gamma$ or to $Z^0 \gamma$ at LEP II.  In either
case, the decay would be easy to identify because of the monochromatic
photon in the final state.

\section{Conclusions} \label{sec:conc}

We have considered the detection of the technirho in technicolor
models that include a weak scalar doublet.  Although our arguments
are somewhat crude, we have arrived at the following qualitative
conclusions:

(1)  When the technirho is light, and both technifermions
have large current masses,  it is likely that the mass of the
technirho will be less than twice the mass of the physical technipions
in the low energy theory.  With the two-pion decay mode kinematically
forbidden, the technirho is extremely narrow and the branching fraction
to $e^+e^-$ or $\mu^+\mu^-$ is large enough to allow detection at
$pp$ colliders.  This result holds even when the technirho decays
primarily to $\pi_p^+ W^-$ or $\eta'_T \gamma$.

(2)  When $\delta=1$, and the technirho is the vector
bound state of the the lighter of the two techniquarks, then the
technirho can be light enough to be seen at LEP II.  In this case,
one could look for the decay to $Z \gamma$ if the scalar bound state
$\tilde{\pi}$ is very close to the mass of the $\tilde{\rho}$, or for the
decay to $\tilde{\pi} \gamma$ if the $\tilde{\pi}$ is light. These decay
modes would be easy to identify because of the monochromatic photon
in the final state.

In either case, the copious production of the light technirhos that we have
considered implies that the signals should be easy to see above background.

\begin{center}
{\bf Acknowledgments}
\end{center}

We are grateful to Howard Georgi and Mike Dugan for interesting comments.
We thank Tom Dignan for providing some useful FORTRAN subroutines.
{\it This work was supported in part by the National Science Foundation,
under grant PHY-9218167, and by the Texas National Research Laboratory
Commission, under grant RGFY93-278B.}

\appendix
\section{Appendix: Decay Widths}

In this appendix, we provide the decay amplitudes used to generate
the numerical results presented in the text.  For $\delta=0$,
the amplitude to $l^+l^-$ or $q\overline{q}$ is determined by vector
dominance, through the operator (\ref{eq:c1}). We find
\beq
\Gamma (\rho_T\rightarrow f \overline{f}) = c_1^2 \frac{\alpha^2}{12\pi}
M_{\rho_T} (a^2 + b^2)
\eeq
where
\beq
a = Q + \frac{c^2-s^2}{4c^2s^2}\frac{M^2_{\rho_T}}{M^2_{\rho_T}-M^2_Z} C_V
\eeq
\beq
b = \frac{c^2-s^2}{4c^2s^2}\frac{M^2_{\rho_T}}{M^2_{\rho_T}-M^2_Z} C_A
\eeq
In these expressions, $Q$ is the charge of the fermion in the final
state (measured in units of $e$), while $C_V$ and $C_A$ are its vector
and axial vector couplings to the $Z^0$ (measured in units of
$e/2cs$).  For $f= e$, $u$, $d$, and $\nu_e$, the $(Q,C_V,C_A)$
values are given by (-1,-0.035,-0.5), (2/3,0.19,0.5),(-1/3,-0.345,-0.5)
and (0,0.5,0.5) respectively.

We estimate the decay width to $W^+ W^-$ from the operator (\ref{eq:c5}).
We find
\beq
\Gamma (\rho_T\rightarrow W^+W^-) = c_5^2 \frac{\alpha^2}{384 \pi s^4}
M_W \left[1 + \left(\frac{M_{\rho_T}}{2 M_W}\right)^2\right]
\left[\left(\frac{M_{\rho_T}}{2 M_W}\right)^2 -1\right]^\frac{3}{2}
\eeq
We estimate the decays to $\eta'_T \gamma$ and $\eta'_T Z^0$ from
the operator (\ref{eq:c3}).  We find
\beq
\Gamma (\rho_T \rightarrow \eta'_T \gamma) = c_3^2 \frac{\alpha}{6}
\frac{M_{\rho_T}^3}{(4 \pi f)^2} \left[1- x_0^2\right]^3
\label{eq:eg}
\eeq
\beq
\Gamma (\rho_T \rightarrow \eta'_T Z^0) = c_3^2 \frac{\alpha}{6}
\frac{M_\rho^3}{(4 \pi f)^2} \frac{(c^2-s^2)^2}{4c^2s^2}
\left[1+x_0^4+y_0^4-2x_0^2-2y_0^2-2x_0^2y_0^2\right]^\frac{3}{2}
\label{eq:ez}
\eeq
where $x_0 = M_{\eta'_T} / M_{\rho_T}$ and $y_0 = M_Z/M_{\rho_T}$.
Finally, we estimate $\rho^0_T \rightarrow \pi^+ W^-$ from the
operator (\ref{eq:c5}).  We find
\beq
\Gamma (\rho_T \rightarrow \pi^+ W^-) = c_5^2 \frac{\alpha}{768 s^2}
F M_{\rho_T} \left[1+x^4+y^4-2x^2-2y^2-2x^2y^2\right]^\frac{1}{2}
\eeq
where $x=M_{\pi_p}/M_{\rho_T}$, $y=M_W/M_{\rho_T}$, and where $F$ is given
by
\[
F = \frac{1}{(4 \pi f)^2 M_W^2 M_{\rho_T}^2} \left[
M_{\rho_T}^6 - 2 M_{\rho_T}^4 M_{\pi_p}^2 -3M_{\rho_T}^2 M_W^4
+6M_{\rho_T}^2 M_W^2 M_{\pi_p}^2 \right.
\]
\beq
\left. +M_{\rho_T}^2 M_{\pi_p}^4
+ 2M_{\pi_p}^4 M_W^2 -4 M_{\pi_p}^2 M_W^4 + 2 M_W^6 \right]
\eeq

In the case where $\delta = 1$, the decay widths to $f \overline{f}$,
and $W^+ W^-$ are given by $\frac{1}{2}$  times the expressions
provided above. The decays to $\tilde{\pi} \gamma$ and
$\tilde{\pi} Z^0$ are of the same form as (\ref{eq:eg}) and
(\ref{eq:ez}), with the replacement $M_{\eta'_T}\rightarrow
M_{\tilde{\pi}}$.  The decay widths in the $\delta=1$ case that we have
not already calculated  are those to $Z^0 \gamma$ and to $Z^0 Z^0$
We estimate both from the operator (\ref{eq:c3}) and find
\beq
\Gamma (\tilde{\rho}\rightarrow Z^0 \gamma) = c_3^2 \frac{\alpha^2}{48 \pi}
\frac{M_{\tilde{\rho}}}{4c^2s^2}
\left[1 + \frac{1}{z^2}\right]\left[1-z^2\right]^3
\eeq
\beq
\Gamma (\tilde{\rho}\rightarrow Z^0 Z^0) = c_3^2 \frac{\alpha^2}{24 \pi}
\frac{(c^2-s^2)^2}{c^4s^4} M_{\tilde{\rho}} z^3
\left[\frac{1}{4z^2}-1\right]^\frac{5}{2}
\eeq
where $z=M_Z/M_{\tilde{\rho}}$.


\newpage
\begin{center}
{\bf Figure Captions}
\end{center}

Fig.1.  Vector meson masses $M_V$ in QCD, as a function of
the quark current mass $M_q$.  Masses are in GeV.

Fig.2. Decay width $\Gamma$ in GeV for $\rho^0_T \rightarrow \pi_p^+ W^-$.

Fig.3. Decay widths $\Gamma$ in GeV
for $\rho^0_T \rightarrow \eta'_T \gamma$ and
$\eta'_T Z^0$.

Fig.4. Decay width $\Gamma$ in GeV
for $\tilde{\rho}\rightarrow \tilde{\pi}\gamma$.


\input prepictex
\input pictex
\input postpictex

\newpage
\begin{figure}[p]
\hbox to \hsize{\hfil
\beginpicture
\setcoordinatesystem units < 0.70in, 0.30in>
\setplotarea x from 0.0 to 5.2, y from 0.0 to 10.0
\axis top ticks unlabeled short quantity 27 /
\axis bottom label {$\displaystyle{M_Q}$}
 ticks numbered from 0.0 to 5.2 by 1.0
  unlabeled short quantity 27 /
\axis left label {$\displaystyle{M_V}$}
 ticks numbered from 0.0 to 10.0 by 2.0
  unlabeled short quantity 26 /
\axis right
 ticks
  unlabeled short quantity 26 /
\setlinear
\plot
0.01      0.669
5.00      9.406
/

\put {$\diamond$} at 0.005    0.770
\put {$\diamond$} at 0.2      1.020
\put {$\diamond$} at 1.5      3.097
\put {$\diamond$} at 5.0      9.460

\put {$\rho$} at 0.1 0.4
\put {$\phi$} at 0.4 0.9
\put {$J/\Psi$} at 1.8 2.7
\put {$\Upsilon (1S)$} at 4.4 9.5

\endpicture
\hfil}
\caption{}
\label{fig:one}
\end{figure}
\begin{figure}[p]
\hbox to \hsize{\hfil
\beginpicture
\setcoordinatesystem units < 7.28in, 7.5in>
\setplotarea x from 0.5 to 1.0, y from 0.0 to 0.4
\axis top label {$\displaystyle{M_{\rho_T}=250}$ GeV}
  ticks unlabeled short quantity 26 /
\axis bottom label {$\displaystyle{M_{\pi_p}/M_{\rho_T}}$}
 ticks numbered from 0.5 to 1.0 by 0.1
  unlabeled short quantity 26 /
\axis left label {$\displaystyle{\Gamma}$}
 ticks numbered from 0.0 to 0.4 by 0.1
  unlabeled short quantity 21 /
\axis right
 ticks
  unlabeled short quantity 21 /

\plot
   .50    .390045
   .51    .374820
   .52    .359510
   .53    .344105
   .54    .328591
   .55    .312944
   .56    .297138
   .57    .281132
   .58    .264876
   .59    .248300
   .60    .231313
   .61    .213787
   .62    .195541
   .63    .176312
   .64    .155681
   .65    .132936
   .66    .106645
   .67    .0729462
   .68     0.00
/

\put {$\rho^0_T \rightarrow \pi_p^+ W^-$} at 0.80 0.20
\endpicture
\hfil}
\caption{}
\label{fig:piw}
\end{figure}
\begin{figure}[p]
\hbox to \hsize{\hfil
\beginpicture
\setcoordinatesystem units < 7.28in, 7.5in>
\setplotarea x from 0.5 to 1.0, y from 0.0 to 0.4
\axis top label {$\displaystyle{M_{\rho_T}=250}$ GeV}
  ticks unlabeled short quantity 26 /
\axis bottom label {$\displaystyle{M_{\eta_T'}/M_{\rho_T}}$}
 ticks numbered from 0.5 to 1.0 by 0.1
  unlabeled short quantity 26 /
\axis left label {$\displaystyle{\Gamma}$}
 ticks numbered from 0.0 to 0.4 by 0.1
  unlabeled short quantity 21 /
\axis right
 ticks
  unlabeled short quantity 21 /

\plot
    .500000     .325336
    .510000     .312369
    .520000     .299505
    .530000     .286759
    .540000     .274148
    .550000     .261687
    .560000     .249391
    .570000     .237276
    .580000     .225355
    .590000     .213645
    .600000     .202157
    .610000     .190906
    .620000     .179906
    .630000     .169167
    .640000     .158704
    .650000     .148527
    .660000     .138647
    .670000     .129074
    .680000     .119819
    .690000     .110890
    .700000     .102296
    .710000     .0940440
    .720000     .0861407
    .730000     .0785920
    .740000     .0714030
    .750000     .0645778
    .760000     .0581193
    .770000     .0520298
    .780000     .0463102
    .790000     .0409606
    .800000     .0359796
    .810000     .0313650
    .820000     .0271132
    .830000     .0232193
    .840000     .0196772
    .850000     .0164793
    .860000     .0136167
    .870000     .0110791
    .880000     .00885454
    .890000     .00692967
    .900000     .00528944
    .910000     .00391721
    .920000     .00279462
    .930000     .00190158
    .940000     .00121621
    .950000     .00071476
    .960000     .00037162
    .970000     .00015919
    .980000     .00004789
    .990000     .00000608
/
\plot
    .500000     .0381201
    .510000     .0341001
    .520000     .0302022
    .530000     .0264390
    .540000     .0228240
    .550000     .0193716
    .560000     .0160977
    .570000     .0130199
    .580000     .0101585
    .590000     .00753766
    .600000     .00518727
    .610000     .00314775
    .620000     .00148014
    .630000     .000301966
    .640000     .00000
/

\put {$\rho^0_T \rightarrow \eta_T' \gamma$} at 0.75 0.15
\put {$\rho^0_T \rightarrow \eta_T' Z^0$} at 0.60 0.08
\endpicture
\hfil}
\caption{}
\label{fig:etap}
\end{figure}
\begin{figure}[p]
\hbox to \hsize{\hfil
\beginpicture
\setcoordinatesystem units < 7.28in, 24.0in>
\setplotarea x from 0.5 to 1.0, y from 0.0 to 0.125
\axis top label {$\displaystyle{M_{\tilde{\rho}}=180}$ GeV}
  ticks unlabeled short quantity 26 /
\axis bottom label {$\displaystyle{M_{\tilde{\pi}}/M_{\tilde{\rho}}}$}
 ticks numbered from 0.5 to 1.0 by 0.1
  unlabeled short quantity 26 /
\axis left label {$\displaystyle{\Gamma}$}
 ticks numbered from 0.0 to 0.125 by 0.025
  unlabeled short quantity 26 /
\axis right
 ticks
  unlabeled short quantity 26 /

\plot
    .500000     .121431
    .510000     .116591
    .520000     .111789
    .530000     .107032
    .540000     .102325
    .550000     .0976740
    .560000     .0930847
    .570000     .0885627
    .580000     .0841135
    .590000     .0797424
    .600000     .0754547
    .610000     .0712554
    .620000     .0671494
    .630000     .0631414
    .640000     .0592360
    .650000     .0554373
    .660000     .0517496
    .670000     .0481768
    .680000     .0447223
    .690000     .0413896
    .700000     .0381818
    .710000     .0351017
    .720000     .0321518
    .730000     .0293343
    .740000     .0266510
    .750000     .0241035
    .760000     .0216929
    .770000     .0194200
    .780000     .0172852
    .790000     .0152884
    .800000     .0134293
    .810000     .0117069
    .820000     .0101199
    .830000     .00866655
    .840000     .00734446
    .850000     .00615085
    .860000     .00508241
    .870000     .00413525
    .880000     .00330494
    .890000     .00258648
    .900000     .00197427
    .910000     .00146209
    .920000     .00104309
    .930000     .000709761
    .940000     .000453947
    .950000     .000266784
    .960000     .000138706
    .970000     .000059417
    .980000     .000017874
    .990000     .000002268
/

\put {$\tilde{\rho} \rightarrow \tilde{\pi} \gamma$} at 0.75 0.06
\endpicture
\hfil}
\caption{}
\label{fig:four}
\end{figure}
\end{document}